# Analysis of network by generalized mutual entropies

V. Gudkov and V. Montealegre

Department of Physics and Astronomy, University of South Carolina, Columbia, SC

29208

**Generalized mutual entropy is defined for networks and applied for analysis of complex network structures. The method is tested for the case of computer simulated scale free networks, random networks, and their mixtures. The possible applications for real network analysis are discussed.**

## 1. Introduction

One of the problems that arise when dealing with complex networks is to understand the network structure and topology in an effective and systematic reliable way. Each type of network can be classified according to different criteria (see, for example [1] [2] [3] [4] and references therein), that can include degree distribution of nodes, cluster structure, the number of giant components present in it, etc. To describe network structure and dynamics in a quantitative way it is desirable to use a function that describes the network features within its range of values. This would provide a way to differentiate different types of complex networks from each other, and even understand the network structure by means of calculating the internal function for clustered sub graphs inside the network.

Also, this could be useful when only partial data of a whole network are available. Thus one could calculate such function on the available set of data to provide a clue about what is the general structure of the network, what types of changes are being made to the evolving network, provided the representative subset of the whole network is chosen. represents.

Using this idea we develop an approach based on information theory to define generalized mutual information of a given network. Then, by relating the entropy to the information in a regular way, we have suggested some hypothesis which has been tested on simulated of networks with a priori defined properties.

We explore the concept of entropy, as a fundamental concept of statistical physics, which is a characteristic of the state of a system and contains information related to general organization of the system such as the level of disorder. Since real networks usually are very large, one can try to describe them in terms of macroscopic parameters in a way similar to the one used in statistical physics. Also we are interested in a function that can describe not only the general (topological) features of the network, but it's interconnectivity as well. Therefore, we define a function that takes into account not only properties of the nodes in the network (e. g., their degree of connectivity), but also parameters of interactions (information exchange) between each pair of nodes. Thus we are able to address the problem of how organized (or disorganized) is the network in general describing it in terms of the entropy and mutual entropy functions. There are other reasons, which are beyond the scope of this paper but should be noted as a guidance, why we would like to explore a generalized entropy approach for description of network dynamics. For example, it has been shown [5], that Rényi entropy can be considered as a measure of localization of complex systems. Another observation is the relation between Rényi entropies and multifractal dimensionalities [6,7] of local sub-structures of complex systems, which may be helpful for the description and understanding of the dynamics and topology of complex networks. However, we do not address these issues in this paper but rather focus on basic features of generalized mutual entropy as a function for description of complex networks.

## 2. Generalized mutual entropy of a network

To define entropy on a network we use a probabilistic approach for information/entropy description. Then, the entropy can be calculated in the usual way, provided the probability distribution function of the network is given. We assume that the network is completely defined if we know the adjacency (connectivity) matrix $C$ as a function of time. Elements of the matrix $C$ contain information about connectivity between nodes and can describe details of the "connections": intensity of connections, details of information exchange, etc. In the simplest case, elements $C_{ij}$ are either equal to 1 (nodes $i$ and $j$ are connected) or 0 (disconnected). However, in general they can be represented by real numbers (intensity of connections) or functions which describe interactions between nodes in the given network. Let us define the "probability" $p_k$ for each node $k$ as

$$p_k = \sum_{i=1}^{n} C_{ik} / \sum_{i,j=1}^{n} C_{ij},  \qquad (1)$$

which, in the simplest case, corresponds to the probability for the node '$k$' to be connected to other nodes in the network.

Now, one can calculate the entropy of the network using Shannon's formula

$$S = -\sum_{k=1}^{n} p_k \log_2 p_k \qquad (2)$$

The Shannon entropy corresponds to Boltzmann entropy in thermodynamics (where $p_k$ is the probability of the system to be found in the $k$-th microstate). However, if we consider a general relation between entropy, as a measure of disorder of a system, and information, as a measure of our knowledge of the system (i.e. entropy is equal to information with an opposite sign), we immediately get to the question: *in how many ways can we define the entropy/information of the system?* The answer for this question is known as Kolmogorov-Nagumo theorem[8][9], which leads to only one other option

$$R_q = \frac{1}{(1-q)} \log\left(\sum_{k=1}^{n} p_k^q\right), \qquad (3)$$

and called Rényi information/entropy of order $q$ ($q \geq 0$). It should be noted that for $q=1$ it coincides with the Shannon entropy/information.

It is worth noting that we are interested in describing the state of the nodes in a network in a general way. This requires a class of functions that provides us not only with information relevant to understand the probabilities of nodes having a particular degree, but also about how the connectivity of one particular node is related to the connectivities of other nodes in the network. Therefore, it is natural to use not only the entropy of the whole network described above but rather the mutual entropy (information)[10] which is defined in terms of conditional information (in a similar way as a conditional probability is defined). Following the standard definition (see, for example [3]), mutual information $I(\xi,\eta)$ (Shannon or Rényi) for two probability distributions $\zeta$ and $\eta$ can be written as

$$I(\xi,\eta) = I(\xi) - I(\xi|\eta) = I(\xi) + I(\eta) - I((\xi,\eta)) \qquad (4)$$

Here $I(\xi)$ is the usual Shannon/Rényi information for $\zeta$-probability distribution, $I(\xi|\eta)$ is the conditional information, and $I((\xi,\eta)) = I(\eta) + I(\xi|\eta)$ is a two dimensional information, which is equal to a sum of information $I(\eta) + I(\xi)$ when $\zeta$ and $\eta$ are independent distributions. It is easy to show [11] that for the above defined network probability functions $P$ mutual information of network can be written as

$$I(C) = I(P(row)) + I(P(column)) - I(P(column)|P(row)), \qquad (5)$$

where for the Shannon case

$$I(P) = \sum_{j=1}^{n} p_i \log p_i \qquad (6)$$

and

$$I(P(column)|P(row)) = \sum_{i,j}^{n} C_{ij} \log(C_{ij}). \qquad (7)$$

For the case of Rényi one has:

$$I_q(P) = -\frac{1}{1-q} \log \sum_{j=1}^{n} (p_i^r)^q \qquad (8)$$

and

$$I_q(P(column)|P(row)) = -\frac{1}{1-q} \log(\sum_{i,j}^{n} C_{ij}^q). \qquad (9)$$

Considering the possibility that elements of the connectivity matrix can be real numbers, we call "simple" or S-type $C$ matrices the connectivity matrices built with only 0's and 1's, on the other hand, if the connectivity matrix contains real numbers, we refer to them as "real" R-type $C$ matrices.

The equations (6) - (9) allow us to find relations between the generalized mutual entropy (using standard equivalence of an entropy as a negative information) of a network and other functions, namely: the (one-dimensional) generalized entropy of node degree distributions $D_q$, and an average correlator $K_q$ whose role is to quantify the divergence of the mutual information function for an R-type network compared to that of an S-type network with the same topological structure. This relation can be expressed as a theorem:

Theorem:

The generalized mutual entropy $H_q(C)$ for the connectivity matrix C can be represented as the sum of the generalized entropy $D_q$ of the degrees of connectivity distribution of the nodes and the average correlator $K_q(C)$:

$$H_q = 2D_q - \log N - K_q(C), \tag{10}$$

where $N$ is the total number of links in the network, $D_q(C) = \dfrac{1}{1-q} \log \sum_{i=1}^{n}(P_i^q)$, and:

$$K_q(C) = \begin{cases} \log \sum_{ij}^{n} \dfrac{C_{ij}^q}{N} & \text{for } q \neq 1 \\[2ex] \dfrac{1}{N}\left(\sum_{ij}^{n} C_{ij} \log C_{ij}\right) & \text{for } q = 1 \end{cases}$$

The interpretation of the theorem is thus immediate: The mutual entropy of a network is the sum of the entropy of the degree distribution, which corresponds to a topological structure (a factor of two comes from symmetry of the *C*-matrix), and the entropy due to "link structure" between nodes, which corresponds to interactions between nodes. For example, for evenly connected nodes (each element in the *C*-matrix has the same link intensity) all correlators $K_q(C)$ are equal to zero. Therefore, all nodes are interacting "equally" and the entropy of interactions (information exchange between nodes) has maximum value $\log N$ (since for an even degree distribution $p_i = 1/N$, and the entropy is equal to $\log N$), which corresponds to a completely disordered state of the system. The increase of the order in information exchange scheme leads to non-zero correlators which decrease the total entropy.

It is obvious that for S-type connectivity matrices (no structure in the information exchange) entropy of the network can be attributed to the particular degree distribution and to the maximum "interaction" entropy $\log N$ (all correlators are zeros). This fact can be stated as a corollary of the theorem above, as follows:

Corollary:

For the S-type connectivity matrix *C* the generalized mutual entropy $H_q(C)$ of the network contains exactly the same information as the (one-dimensional) generalized entropy $D_q$ of the degree of connectivity of nodes. They are related by the equation

$$H_q = 2D_q - \log N \tag{11}$$

The last corollary is of particular importance for structural analysis purposes, since it shows clearly that for a network containing only binary information about the connections between nodes, the bulk of the information is contained in the degree distribution, while for a network with different intensities associated to its links, the part of the mutual entropy can come from the structure of information exchange between nodes (see section 4).

Understanding the meaning of generalized entropies based on the above considerations while taking into account the properties of entropy function in thermodynamics leads to a number of conjectures. In particular we focus our attention on the following two hypotheses: the set of entropies can be used to characterize the main properties related to the structure (topological and information exchange) and dynamics of networks; and that the entropies, calculated over a representative part of the network contain enough information to describe the whole network. To test these conjectures and their possible applications, we need to study networks with different natures (scale-free, random, etc.) and different sizes. It is also important to identify which contributions to the parameters under investigation come from topological network structure, from interactions between nodes, and from random noise which does not change major network characteristics. For this purposes we use simulation and analysis techniques which are described below.

## 3. Network simulation and analysis

**Network Generation Algorithm**

For the current analysis we use two different types of network: scale-free and random networks (see, for example [12], and references therein), as well as a mixture of these two networks with the controlled weights of each one.

To create a "mixed" type of network we followed the procedure suggested in the paper [13], where the network growth model creates each new link using a preference attachment rule according to the probability:

$$\Pi_i = \frac{(1-p)d_i + p}{\sum_{j=1}^{t}\left[(1-p)d_j + p\right]}. \quad (12)$$

Here $d_i$ is a degree of $i$-th node, $p$ is a "weighting" parameter whose values lie in the range between 0 and 1 ($0 \leq p \leq 1$) so a scale-free network is generated for the value $p = 0$, and a random network for $p = 1$, correspondingly.

For our simulation we speeded up the algorithm by using the attachment probability equation

$\Pi_i = w_1 P_{SF} + w_2 P_R$, with $w_1 + w_2 = 1$, $P_{SF} = d_i/D$, $P_R = 1/N$, and $D = \sum_i d_i$; which is numerically equivalent to the algorithm discussed above. Thus, new connections are assigned according to weights $w_1$ and $w_2$ in a "scale-free" or "random" way respectively for $P_{SF}$ and $P_R$.

The efficiency of our computer algorithm is tested by $\chi^2$-method for the linear fits of the degree distributions with the power law and exponential regressions. It shows that the algorithm resembled correctly the theoretical expectations, and by means of a quality factor for the fits, we determined that the minimum acceptable size of the networks to be free from possible systematic errors coming from the simulation algorithm (in order to keep consistency with the theory) is about 3000 nodes. Therefore, to avoid systematic errors in the simulations we worked with networks of the size of 3000 and larger.

**Perturbation of a network: "noise-to-signal ratio"**

Topological structure of a network is mainly defined by the subset of nodes with large degrees of connectivity. To be able to recognize this relatively steady topological structure and filter it from the background of random connections, we use an algorithm that perturbs the network without visible changes of its structure. The main idea implemented in the algorithm is described in the paper [14]: the most relevant links between nodes are reconnected according to a probability which is proportional to the product of the degrees of each couple of nodes in the network. Then the probability of a link to survive in a network after such transformation is

$$p_{ij} = d_i d_j / \sum_{k=1}^{n} d_k, \tag{13}$$

where $d_i$ is the degree of the node $i$ in the network, and $n$ is the total number of nodes. To allow only small perturbations of the network topology, we use a threshold to filter the connections $t = c/(1+\tilde{d})$ where $\tilde{d} = \sum_{i=1}^{n} d_i^2 / \sum_{i=1}^{n} d_i$, and $c$ is a constant. To preserve some of the nodes with smaller connectivity degree that could be accounted for part of the structure of the network, we assign the threshold randomly with a Gaussian distribution. Thus, we found out that by choosing $c = 1$, and the dispersion parameter for Gaussian distribution $\sigma = 0.35$, the perturbated network corresponds accurately to the original one based on the analysis of all involved parameters (degree distribution and entropies).

**Clustering Algorithm**

To study the properties of the mutual entropy of the whole network and its sub networks, it is necessary to be able to identify network structure: all clusters inside the network. There is a number of methods for cluster analysis of networks (see, for example, [15] [16] [17] and references therein). For our purposes we use the physics based model [18], which provides detailed (spectroscopic) information about network structure and clusters substructures at the same time during one run. The main idea of the method is to eliminate $n!$ permutations for the possible network description, in terms of the adjacency matrix, by choosing absolutely symmetric initial conditions: all $n$ nodes are equidistant points over a one ($n$-$1$) dimensional sphere. The links between nodes are simulated by attractive/repulsive forces. Then, the mostly connected nodes (clusters) start gathering (condensate) in groups corresponding to the network communities governed by the connectivity matrix $C$. This algorithm resolves the network structure from a fully shuffled (randomly permuted) connectivity matrix into clearly defined clusters after only a few steps.

# 4. Application of generalized entropy for the analysis of simulated networks

In this paper we simulate networks of the size 3000 nodes or larger to make sure that the network structure does not depend on the initial conditions (as it was discussed in the previous section, we do not see such dependence starting from the size of about 3000 nodes). First we generated a number of networks with different mixing parameter '$p$', (see Eq.(12)) and applied the clustering algorithm to verify that the most densely connected regions of the networks are identified. We observed that they follow a self-similar pattern and the traits of this pattern for mixed networks disappear in a progressive way as the value of the mixing parameter increases ($p = 0.0$, $p = 0.3$, $p = 0.7$, $p = 1.0$ are shown in figures 1a, 1b, 1c and 1d, correspondingly). This change of pattern illustrates that scale-free networks ($p = 0$) possess a more organized structure than random networks ($p = 1$). The pattern of self-similar structure for the scale-free network shows hierarchical connectivity organization of larger clusters containing the smaller ones. As the value of $p$ increases, the sizes of the clusters become smaller, and practically disappears for totally random network at $p = 1$.

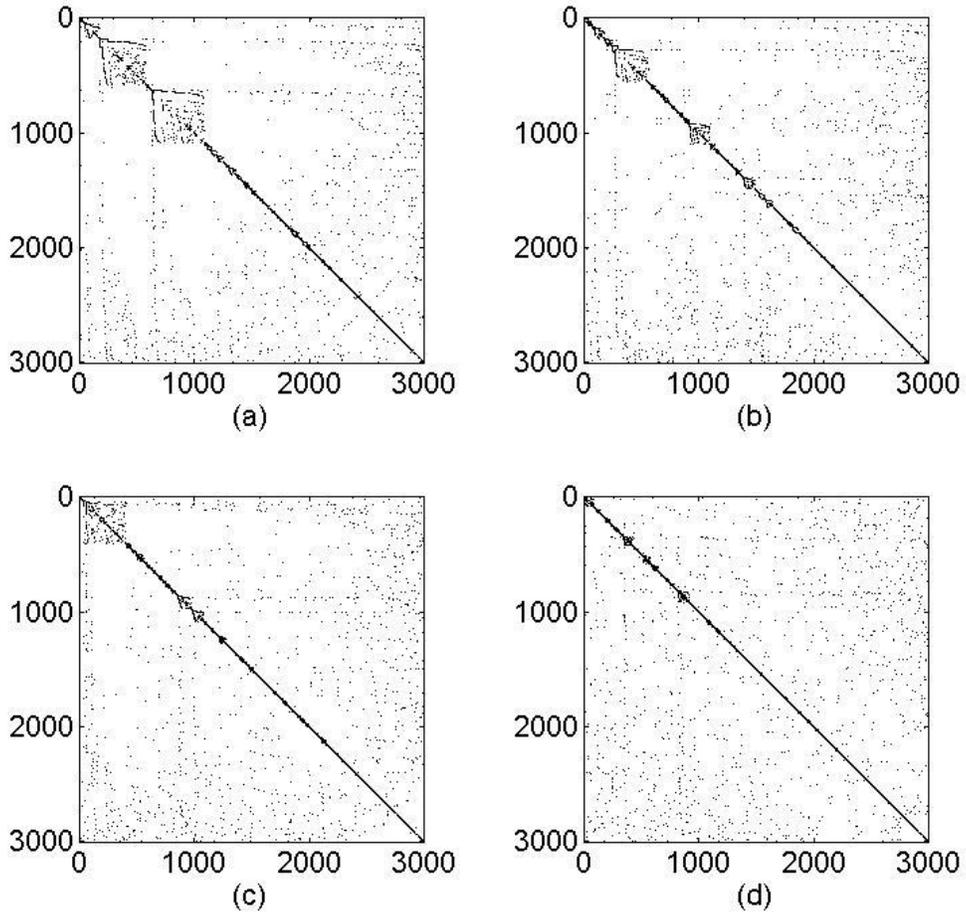

**FIGURE 1.** Connectivity matrices of networks with 3000 nodes for values of the mixing parameter $p$=0.0, 0.3, 0.7 and 1.0 are shown in Fig. 1a, 1b, 1c and 1d respectively. It is observed that the cluster structure disappears as the value of $p$ increases.

Let us focus on the case of scale-free networks ($p$=0) to clarify contributions to the entropies from topological structure of networks and from "information-exchange" between nodes (see the theorem in sec. 2). Figure 2 shows the components of the mutual entropy for a scale-free network of 5000 nodes as functions of order of entropy $q$. To be able to distinguish between topological and "information-exchange" contributions we assign the values of the non-zero elements of the adjacency matrices of the simulated networks using three essentially different prescriptions: a) the matrix elements values $C_{ij}$ are equal to the product of the degrees of nodes $i$ and $j$; b) their values are equal to inverse product of the node's degrees; and c) the values of matrix elements are uniformly distributed as real numbers in the range between 0 and 1; (see figures 2a, 2b and 2c, correspondingly). On these figures the generalized mutual entropies $H$ of the network (shown by circles), generalized entropy D of degree distributions (by squares) and the average correlator K (by stars), are plotted against $q$. Filled symbols represent the values for the S-type matrix, and the hollow symbols the R-type matrix.

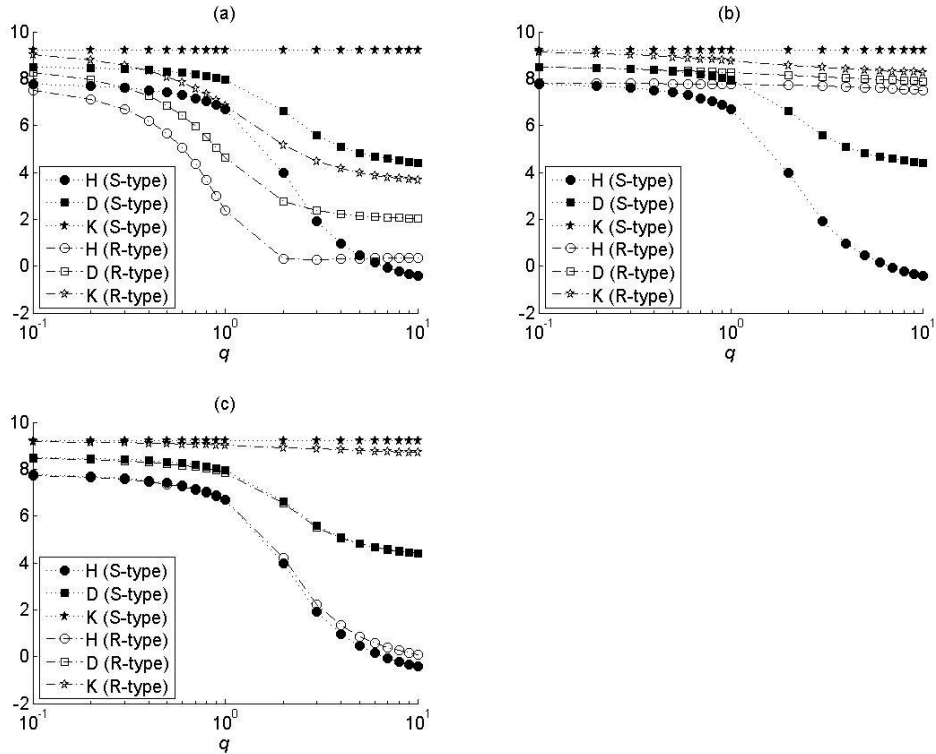

**FIGURE 2:** The generalized mutual entropy H, the generalized entropy D and the average correlator K, are plotted versus the mutual entropy order ($q$) when the nonzero matrix elements of the connectivity matrix are replaced by the products of the nodes degrees (a), the inverses of the products of the nodes degrees (b), and a uniform distribution of real numbers in the range between 0 and 1 (c).

One can see that the mutual entropy for S- and R-types of networks (and their components) differs significantly only when the strengths of the links (values of matrix elements) vary in a large range of values. If the connection strength follows a flat distribution of random numbers the mutual information does not differ very much between R- and S- types. This is because the information-exchange part became practically unstructured (disordered) and, as a consequence, variations of the total entropy are determined mostly by contributions from the topological part. Thus, to determine the relation between network topology and mutual entropy one can use S - type networks.

To see how entropy depends on the size of a sub-network we calculate the mutual entropy for different $q$-orders from randomly selected subsets of nodes of different sizes in a scale-free network. The mutual entropies of order 0, 1, 2 (and the difference between orders 1 and 2) plotted versus the size of randomly chosen sub networks are shown in Figure 3, where circles are used for $q=0$, triangles for $q=1$, squares for $q=2$ and stars for differences of $q=1$ and $q=2$.

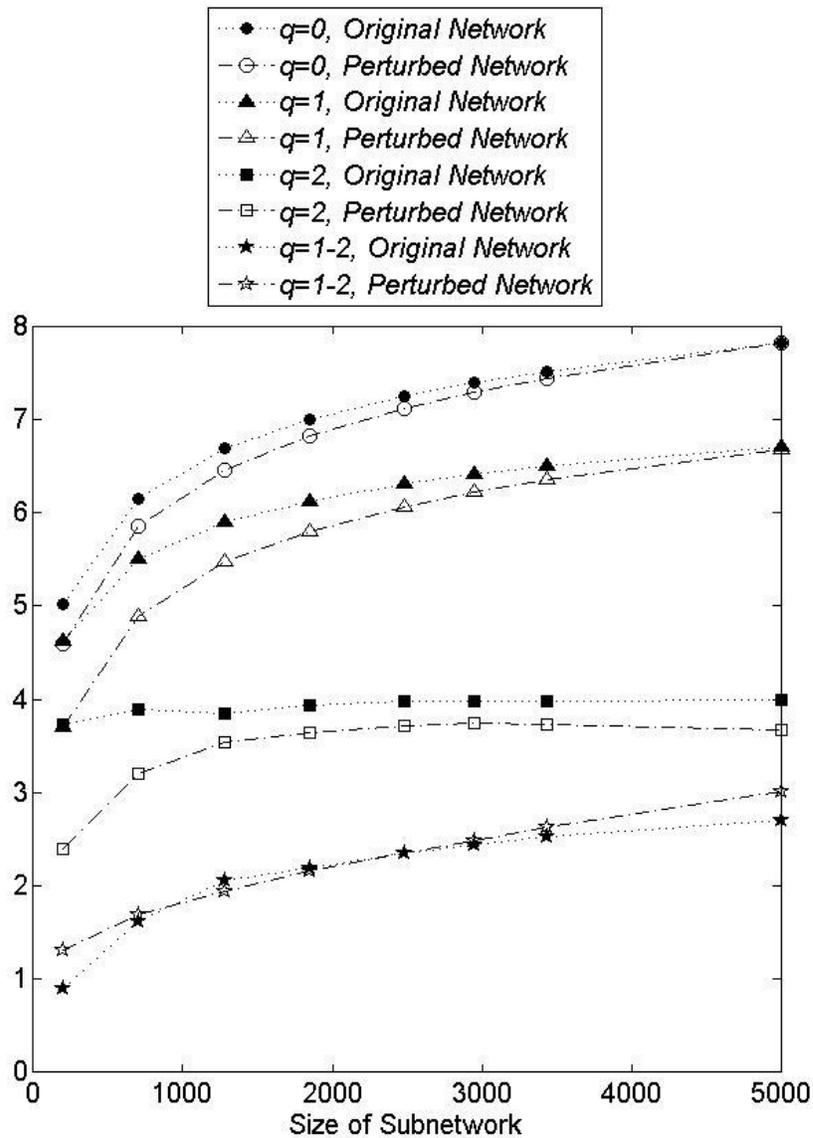

**FIGURE 3.** Mutual Entropies of randomly selected subsets of nodes of a matrix with size 5000 nodes, circles, triangles, squares and stars represent values of $q = 0, 1, 2$ and the difference of the entropies for $q = 1$ and $q = 2$, respectively. Filled symbols are used for the original network, and the hollow symbols are for the perturbed version of it.

The filled circles, triangles, squares and stars indicate mutual entropies calculated for the randomly chosen sub networks as they are originally created; and the hollow ones are for perturbed sub networks. One can see that the differences between the original and the perturbed networks become larger as the size of the sub network decreases, thus reflecting the fact that smaller sub networks contain less information about the whole network. Taking into account that the sub-networks are chosen randomly (without any preference to the particular cluster at the given network), one can consider this as a worst choice in determination of the representative part of a network. However, even this worst case demonstrate that the mutual entropies come to a saturation value for the size of a sub-network by about of the half of the size of the whole network. Therefore, the results on the figure 3 clearly support our conjecture that the entropies of the sub-networks can be used to characterize the whole network. In other words, loss of information about small parts of a network is nor critical if we use entropies as a measure of network characteristics. It should be noted that the saturation behavior of entropies shown on fig. 3 practically does not depend on the nature of network (in our case on the mixture parameter $p$) which indicate the universal property of the mutual entropy as a characteristic of a network.

This result can be used in a number of applications since it does not depend on the nature of network, however one can choose a sub-network selectively, based on a closed (or other preferred) cluster. To study this option we consider the same scale-free network but we identify the cluster (sub-network) structure first, and only after that organize the network according its clusters structure as shown in figure 4.

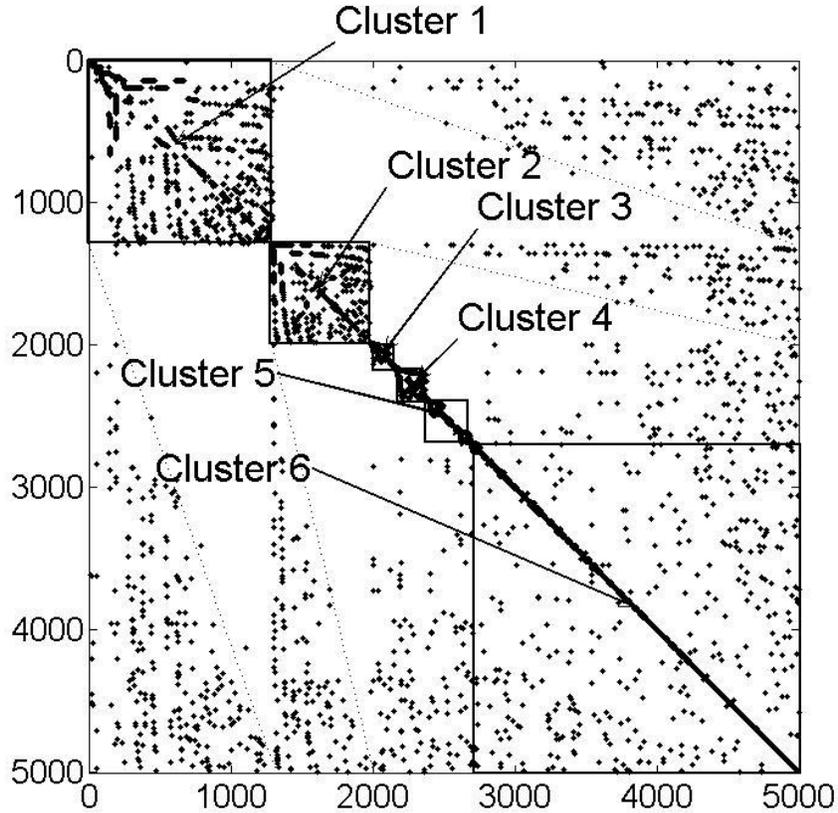

**FIGURE 4.** Six main clusters are identified for a scalefree network of size 5000 nodes showing a pattern of self similarity.

To relate the information we can obtain from a selected cluster to the entropy of the whole network structure, we plot the mutual entropy corresponding to the cropped section of the connectivity matrix that contains the cluster versus its size (see figure 5). There, mutual entropy for clusters in the 5000 nodes scale-free network for order $q=0$ shown in the fig. 5a, $q=1$ in the fig. 5b, $q=2$ in the fig. 5c, and for the difference of mutual entropies with $q=1$ and $q=2$ in the fig. 5d.

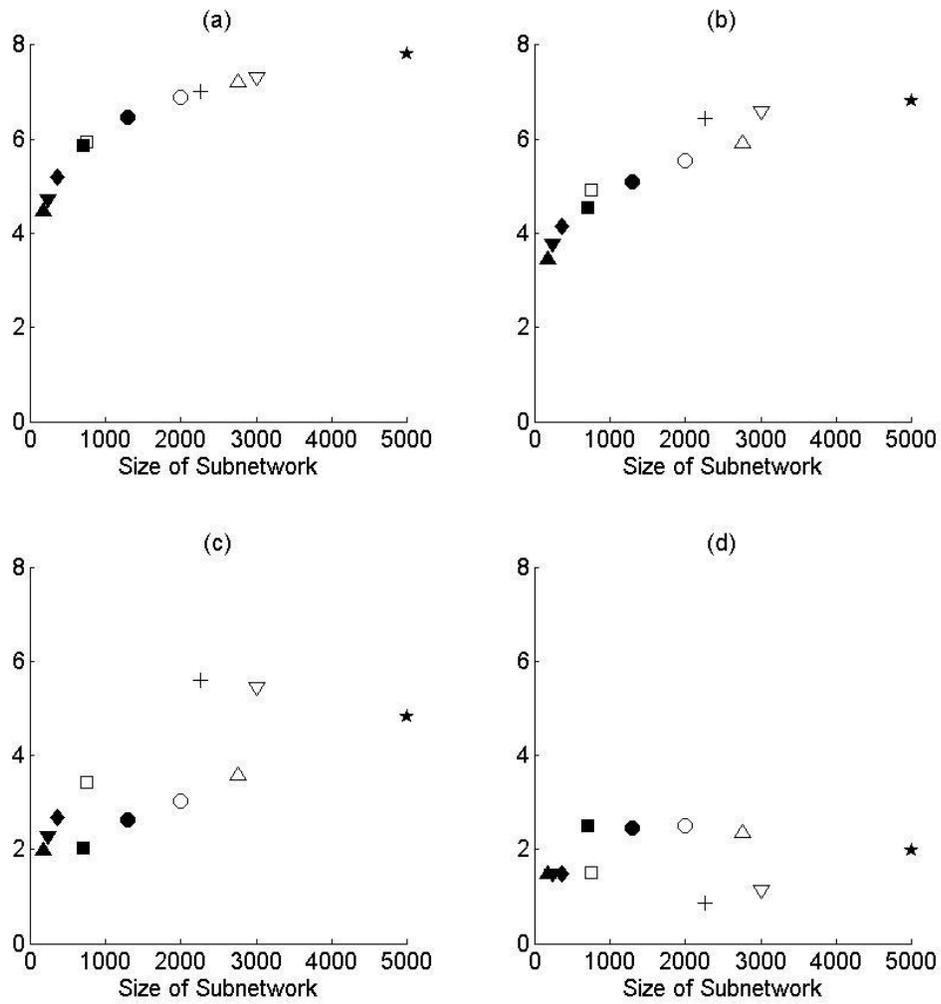

**FIGURE 5.** Mutual entropy for clusters in the 5000 nodes scalefree network. The graphs correspond to $q=0$ (a), $q=1$ (b), $q=2$ (c), and $\Delta q$=1-2 (d)

| **TABLE 1.** Correspondence of the symbols in fig. 5 to the clusters identified in fig. 4 ||
| --- | --- |
| **Clusters contained** | **Symbol used** |
| 1 | Filled Circle |
| 2 | Filled Square |
| 3 | Filled upward-pointing triangle |
| 4 | Filled downward-pointing triangle |
| 5 | Filled Diamond |
| 6 | Plus sign |
| 1 and 2 | Circle |
| 3, 4 and 5 | Square |
| 1, 2, 3, 4, 5 | Upward pointing triangle |
| 3, 4, 5, 6 | Downward pointing triangle |
| Whole Network | Filled star |

These figures show that the mutual entropy of the order 0 (fig. 5a) follows the same pattern that could be expected if the sub networks were chosen by picking up nodes randomly. This is in agreement with the fact that mutual entropy of order zero simply accounts for the number of connections in the network (sub network).

Starting from the order one we observe a cluster differentiation (fig. 5b): two points got off the trend of the sequence. Those two points correspond to cluster 6 and the sub network comprising the nodes in clusters 3, 4, 5 and 6, which have different structures. For the case of order two (fig. 5c) we observe even more accentuated entropy difference for these clusters, and a deviation from the trend for the entropy of clusters 4 and 5. This corresponds to the fact that for lager degree of the mutual entropy the cluster structure differentiation is more refine due to a power-dependence in the degrees of probabilities (see, for example Eq.(3)). It is interesting that the difference of mutual entropies of order 1 and 2 shows a clear flat trend for the scale-free sub networks (in fig. 5d). Their values for clusters 3, 4, 6 and the cluster (3 + 4 + 5 + 6) are separately grouped, clearly showing that the mutual entropy difference is sensitive to the internal structure of clusters and could be used for identification of sub-network structure.

Figure 6 presents mutual entropy dependence on the choice of the sequences of the clusters in a given network.

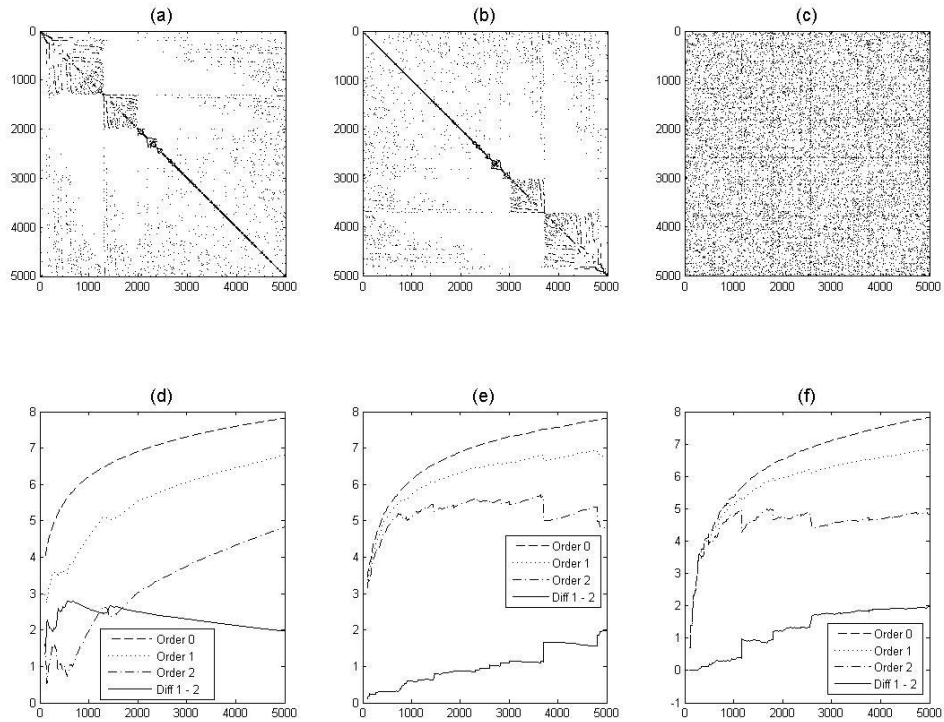

**FIGURE 6.** Changes in the mutual entropy as a function of the size of the sub network with nodes included according to the size of the clusters.

The mutual entropy obtained by accumulating the effects of the larger clusters first and then the most random ones shows rapid fluctuations in the clustered region. This fluctuation is related to contributions from nodes with high degrees in this large cluster.

After the larger cluster has been absorbed, the fluctuations disappear, and the function becomes smooth and monotonically increasing, since the nodes with small degrees increase entropy smoothly by adding disorder into the structure of the system. For the reversely ordered nodes, the mutual entropy is monotonically increasing for orders 0 and 1, and starts showing some fluctuations for order 2. One can see a drop in the mutual entropy of order one just after the first large cluster has been included completely. This fact indicates the possibility to use mutual entropy to estimate a level of organization of a sub-network. For example, mutual entropies can be used as a tool to differentiate a random sub-network from a structured one. For the completeness of the consideration we present the mutual entropy dependence for a random choice of the sub-network in fig. 6f (see corresponding nod's order on the fig. 6c), which was discussed in detail with relation to the fig. 3.

## Conclusions

We have studied basic properties of generalized mutual entropy of a network using different types of simulated networks. It was shown, by calculating the Rényi entropy for identifiable clusters of nodes within a given network, that this approach gives the opportunity to differentiate network substructures. The results of this paper demonstrate that scale free networks possess a hierarchical structure that mimics itself in its main building blocks which can be identified by means of comparing the whole network's mutual entropy to that of a perturbed version of it. Thus our method can be used to identify the most sensitive groups of nodes that make a scale free network more vulnerable because they contain most information about the global network structure which can be extracted from a selected representative part of the whole network. Moreover, the network description provided by the mutual entropy can be used as a measure of the level of organization in network's structure after being modified by a perturbation and to indicate which part of the network has been changed. Therefore, this is a promising tool to study the network's evolution. We have also shown that the analysis of different $q$-degree entropies of the network is indeed an efficient measure to distinguish the contribution of each sub network to the global network structure, both to the topology and to information exchanges between nodes. This is achievable because we use not only Shannon's entropy but the whole set of possible entropies (in general, the infinite set of Rényi entropies, defined for any positive $q$) each of which is sensitive to specific properties of the network. We have proved numerically that one can make conclusions about structure/dynamics of a whole network by analyzing only a representative part of it. This may lead to a promising method of analysis of real networks, when the number of nodes unknown and/or changes with time.

Our analysis has been focused on scale free networks, random networks and mixed networks of those two kinds. We plan to apply this procedure for analysis of networks with different topological structures of the same type (hubs, trees, rings, etc) as well as to networks composed of sub networks of different types.

## Acknowledgments

This work was supported by DARPA through AFRL grant FA8750-04-2-0260.